\documentclass[aps, pre, twocolumn, superscriptaddress]{revtex4-1}

\usepackage{amsfonts, amsmath, amssymb, bbold, bigints, color, enumitem, epstopdf, float, graphicx, lineno, makeidx, rotating, setspace, subfigure, url}
\usepackage[section]{placeins}
\usepackage[bookmarks, breaklinks, colorlinks]{hyperref}
\hypersetup{citecolor = blue, filecolor = black, linkcolor = red, urlcolor = blue}
\hypersetup{pdfauthor = {Chiranjit Mitra}}

\begin{document}

\title{Multi-node basin stability in complex dynamical networks}

\author{Chiranjit Mitra}
\email[Corresponding author: ]{chiranjit.mitra@pik-potsdam.de}
\affiliation{Potsdam Institute for Climate Impact Research, Research Domain IV- Transdisciplinary Concepts \& Methods, 14412 Potsdam, Germany}
\affiliation{Humboldt University of Berlin, Department of Physics, 12489 Berlin, Germany}

\author{Anshul Choudhary}
\affiliation{Indian Institute of Science Education and Research (IISER) Mohali, Knowledge City, SAS Nagar, Sector 81, Manauli PO 140 306, Punjab, India}
\thanks{Now at: Theoretical Physics/Complex Systems, ICBM, Carl von Ossietzky University of Oldenburg, Carl-von-Ossietzky-Strasse 9-11, Box 2503, 26111 Oldenburg, Germany}

\author{Sudeshna Sinha}
\affiliation{Indian Institute of Science Education and Research (IISER) Mohali, Knowledge City, SAS Nagar, Sector 81, Manauli PO 140 306, Punjab, India}

\author{J\"{u}rgen Kurths}
\affiliation{Potsdam Institute for Climate Impact Research, Research Domain IV- Transdisciplinary Concepts \& Methods, 14412 Potsdam, Germany}
\affiliation{Humboldt University of Berlin, Department of Physics, 12489 Berlin, Germany}
\affiliation{University of Aberdeen, Institute for Complex Systems and Mathematical Biology, Aberdeen, AB24 3UE, United Kingdom}
\affiliation{Nizhny Novgorod State University, Department of Control Theory, Nizhny Novgorod, 606950, Russia}

\author{Reik V. Donner}
\affiliation{Potsdam Institute for Climate Impact Research, Research Domain IV- Transdisciplinary Concepts \& Methods, 14412 Potsdam, Germany}

\begin{abstract}
Dynamical entities interacting with each other on complex networks often exhibit multistability. The stability of a desired steady regime (e.g., a synchronized state) to large perturbations is critical in the operation of many real-world networked dynamical systems such as ecosystems, power grids, the human brain, etc. This necessitates the development of appropriate quantifiers of stability of multiple stable states of such systems. Motivated by the concept of basin stability (BS) (Menck et al., Nature Physics 9, 89 (2013)), we propose here the general framework of \emph{multi-node basin stability} for gauging global stability and robustness of networked dynamical systems in response to non-local perturbations simultaneously affecting multiple nodes of a system. The framework of multi-node BS provides an estimate of the critical number of nodes which when simultaneously perturbed, significantly reduces the capacity of the system to return to the desired stable state. Further, this methodology can be applied to estimate the minimum number of nodes of the network to be controlled or safeguarded from external perturbations to ensure proper operation of the system. Multi-node BS can also be utilized for probing the influence of spatially localised perturbations or targeted attacks to specific parts of a network. We demonstrate the potential of multi-node BS in assessing the stability of the synchronized state in a deterministic scale-free network of R\"{o}ssler oscillators and a conceptual model of the power grid of the United Kingdom with second-order Kuramoto-type nodal dynamics.
\end{abstract}


\maketitle


\section{\label{sec:Introduction}Introduction}

Multistable dynamical systems are abundant across several disciplines of natural sciences and engineering~\cite{feudel2008complex, pisarchik2014control}. The human brain~\cite{kelso2012multistability}, lasers~\cite{arecchi1982experimental}, ecosystems~\cite{may1977thresholds}, ice sheets~\cite{robinson2012multistability}, delayed feedback systems~\cite{mitra2014dynamical}, chemical reactions~\cite{ganapathisubramanian1984bistability}, any many others constitute notable examples among a large body of multistable systems~\cite{feudel2008complex, pisarchik2014control}. Subsequently, the pervasiveness of multistability in dynamical systems calls for suitable quantifiers of the respective stability of the multiple attractors of such systems.

Linear stability analysis, based on the `local' assessment of the sign and magnitude of the largest Lyapunov exponent in the attractor's neighbourhood is primarily employed in stability assessments of such complex dynamical systems~\cite{strogatz2014nonlinear}. However, linear stability analysis can only assess the vulnerability of the state against small perturbations and subsequently classify the state as stable or unstable. On the other hand, many dynamical systems such as the ocean circulation~\cite{dijkstra2005nonlinear} or the synchronized dynamics in power systems~\cite{machowski2011power} are prone to large perturbations. In this context, the method of constructing Lyapunov functions for determining the stability of equilibria and estimating the basins of attraction was a major theoretical advancement but practically often not applicable~\cite{strogatz2014nonlinear}. Subsequently, there has been a persistent drive towards quantifying the stability and resilience of the multiple (stable) states of such systems against large perturbations~\cite{menck2013basin, menck2014dead, mitra2015integrative, klinshov2015stability, hellmann2016survivability, kittel2016timing, kohar2014synchronization}.

A major advancement in this direction was the development of basin stability (BS)~\cite{menck2013basin, menck2014dead}. The BS of a particular stable state relates the volume of its basin of attraction to the likelihood of returning to the stable state in the face of random perturbations. Formally, the BS of any given attractor $\mathcal{A}$ of a multistable dynamical system (represented by the state vector $\mathbf{x}$) is defined as~\cite{menck2013basin, menck2014dead}
\begin{equation} \label{eq:BS}
S_{B} \left( \mathcal{A} \right) = \int \chi_{\mathcal{B} \left( \mathcal{A} \right)} \left( \mathbf{x} \right)\, \rho \left( \mathbf{x} \right)\, d\mathbf{x},
\end{equation}
where $\chi_{\mathcal{B} \left( \mathcal{A} \right)} \left( \mathbf{x} \right) = 1$ if the state $\mathbf{x}$ belongs to the basin of attraction $\mathcal{B} \left( \mathcal{A} \right)$ of the attractor $\mathcal{A}$ and $\chi_{\mathcal{B} \left( \mathcal{A} \right)} \left( \mathbf{x} \right) = 0$ otherwise. $\rho \left( \mathbf{x} \right)$ is the density of states in state space that the system may be pushed to via large perturbations, with $\int\limits_{\mathbf{x}} \rho \left( \mathbf{x} \right)\, d\mathbf{x} = 1$, where the integral is taken over the entire state space. In order to avoid terminological confusion, we emphasize that $\rho$ is not the invariant density of the attractors.

Many complex systems involve large collections of dynamical units interacting with each other on complex networks~\cite{strogatz2001exploring, newman2003structure}. Coupled map lattices constitute the simplest classes of such systems displaying multistability on account of formation of clusters~\cite{kaneko1990clustering}. Other important examples include coupled weakly dissipative systems, logistic maps, H\'{e}non maps, genetic elements, or mutually coupled semiconductor lasers (cf.~\cite{feudel2008complex} and references therein). Such coupled dynamical systems exhibit a great variety of emergent phenomena with synchronization being the most intensively reported and practically relevant one. In fact, the ubiquity of synchronization in networked dynamical systems can hardly be further exaggerated and plays a central role across various disciplines such as biology, ecology, climatology, sociology, engineering, etc.~\cite{pikovsky2003synchronization, arenas2008synchronization}. The coexistence of synchronized and desynchronized dynamics in such systems is a typical case of bistability. In this regard, the presence of the fully synchronized state (for homogeneous initial conditions) and the chimera state (for particular heterogeneous initial conditions) in networks of oscillators with non-local coupling has gathered a lot of recent attention~\cite{tinsley2012chimera, omelchenko2015robustness}. It is essential to appropriately assess and quantify multistability, particularly, the robustness of the synchronized state to arbitrary perturbations of such coupled dynamical systems. In this direction, the framework of master stability function (MSF)~\cite{pecora1998master} as an extension of the linear stability concept to assess the stability of the completely synchronized state in coupled networks was a considerable development but still locally restrictive to small perturbations. The application of BS to assessing the stability of synchronized dynamics and its extension to the concept of single-node BS in quantifying the contributions of individual nodes to the overall stability of the synchronized state has been a major advancement and complements linear stability analysis substantially~\cite{menck2013basin, menck2014dead}.

Single-node BS of a node under investigation corresponds to the probability of the system (operating in the desired stable state) to return to its stable state after that particular node has been hit by a non-local perturbation~\cite{menck2014dead}. We reserve a formal definition of single-node BS to Section \ref{sec:SNBS}. However, in many practical situations, disturbances affect a group of nodes of the network, significantly hampering its return to the desired operational state. Some of the most relevant examples are the collapse of ecological networks due to spatial perturbations~\cite{pascual2005ecological}, cascading failures in a power transmission grid on account of breakdown of few nodes~\cite{motter2002cascade}, or epileptic seizures triggered by random perturbations of neural networks~\cite{jiruska2014modern}. Subsequently, it is essential to develop a framework for assessing the robustness of networked dynamical systems to withstand perturbations simultaneously hitting several nodes of the system. In addition, such a framework should provide a critical number of nodes which when simultaneously perturbed significantly reduce the probability of the system to continue operating in the desired regime. Further, such a methodology could also solve the associated problem of estimating the minimum number of nodes of the network which need to be safeguarded from external perturbations to ensure proper functionality of the system. As a crucial first step in this direction, we extend here the concept of single-node BS to the general framework of \emph{multi-node basin stability}. We provide a formal definition of multi-node BS in Section \ref{sec:MNBS}.

Previous studies on the robustness of complex networks have mainly focused on static (topological) properties of networks and their ability to withstand failures and perturbations on account of removal of nodes and/or links~\cite{motter2002cascade}. In this context, the framework of percolation theory has generated important insights useful for the analysis and prediction of resilience of complex networks by deriving a critical threshold for the fraction of nodes that need to be removed for the breakdown of the giant component of a complex network~\cite{gao2015recent, cohen2000resilience}. Recently, Gao et al.~\cite{gao2016universal} considered intrinsic nodal dynamics in developing an analytical framework of a universal resilience function to accurately unveil the resilience of networked dynamical systems. However, their approach considers node, weight and link losses as possible perturbations to the system and not actual non-local disturbances to the dynamical state of individual or several nodes of the system, as addressed by single-node BS and multi-node BS, respectively. Also, almost all stability studies assume no knowledge about the nature of perturbations to the system. The framework of multi-node BS developed here can be applied to probe the influence of spatially localised perturbations or targeted attacks to specific parts of a network, which can be practically more relevant.

This paper is further organized as follows: In Section \ref{sec:Methods}, we outline the general methodology for calculating multi-node BS for a given networked dynamical system. In Section \ref{sec:Examples}, we illustrate its application to a deterministic scale-free network of R\"{o}ssler oscillators and a conceptual model of the power grid of the United Kingdom with second-order Kuramoto-type nodal dynamics. Finally, we present the conclusions of our work in Section \ref{sec:Conclusion}.


\section{\label{sec:Methods}Methods}

BS of any particular attractor of a multistable dynamical system is estimated using a numerical Monte-Carlo procedure by drawing random initial states from a chosen `reference subset'~\cite{menck2013basin} of the entire state space, simulating the associated trajectories, and calculating the fraction of trajectories that asymptotically approach the respective attractor. We refer to Menck et al.~\cite{menck2013basin} for further details on the procedure for estimating BS. In the following, we outline the general methodology for estimating single-node BS and multi-node BS values for any networked dynamical system.

Consider a network of $N$ oscillators (nodes) where the intrinsic dynamics of the $i\textsuperscript{th}$ oscillator (represented by the $d$-dimensional state vector $\mathbf{x}^{i}(t) = \left( x_{1}^{i},\, x_{2}^{i},\, \ldots,\, x_{d}^{i} \right)^{\text{T}}$) is described by
\begin{widetext}
\begin{equation} \label{eq:DE_Individual}
\dot{\mathbf{x}}^{i} = \mathbf{F}^{i} \left( \mathbf{x}^{i} \right);\, \mathbf{x}^{i} \in \mathbb{R}^{d};\, \mathbf{F}^{i}:\, \mathbb{R}^{d}\, \rightarrow\, \mathbb{R}^{d},\, \mathbf{F}^{i} = \left( F_{1}^{i} \left( \mathbf{x} \right),\, F_{2}^{i} \left( \mathbf{x} \right),\, \ldots,\, F_{d}^{i} \left( \mathbf{x} \right) \right)^{\text{T}};\, i = 1,\, 2\, \ldots\, N.
\end{equation}
\end{widetext}
The dynamical equations of the networked system read
\begin{equation} \label{eq:DE_Network}
\dot{\mathbf{x}}^{i} = \mathbf{F}^{i} \left( \mathbf{x}^{i} \right) + \epsilon \sum\limits_{j = 1}^{N} A_{ij} \mathbf{H}^{ij} \left( \mathbf{x}^{i},\, \mathbf{x}^{j} \right),
\end{equation}
where $\epsilon$ is the overall coupling strength, $\mathbf{A}$ is the (directed) adjacency matrix which captures the interactions between the nodes such that $A_{ij} \neq 0$ if node $j$ influences node $i$ and $\mathbf{H}^{ij}:\, \left( \mathbb{R}^{d},\, \mathbb{R}^{d} \right)\, \rightarrow\, \mathbb{R}^{d}$ is an arbitrary coupling function from node $j$ to node $i$ such that $\mathbf{H}^{ij}$ and $\mathbf{H}^{ji}$ may be different, in general. For the illustrations in this paper (Section \ref{sec:Examples}), we consider identical nodal dynamics $\left( \mathbf{F}^{i} \equiv \mathbf{F}\, \forall\, i \right)$, symmetric adjacency matrices ($A_{ij} = A_{ji} = 1$ if nodes $i$ and $j$ are connected and $A_{ij} = A_{ji} = 0$ otherwise) and identical coupling functions ($\mathbf{H}^{ij} \equiv \mathbf{H}\, \forall\, i,\, j$). We present the framework (and associated illustrations) for networks of oscillators with continuous time dynamics (Eq.~(\ref{eq:DE_Network})) exhibiting bistability on account of coexisting synchronized and desynchronized regimes. However, the framework is generally applicable to any networked (continuous or discrete time) dynamical system with multiple coexisting states.


\subsection{\label{sec:SNBS}Single-node basin stability}

\begin{figure}
\begin{center}
\includegraphics[height=6.0cm, width=8.0cm]{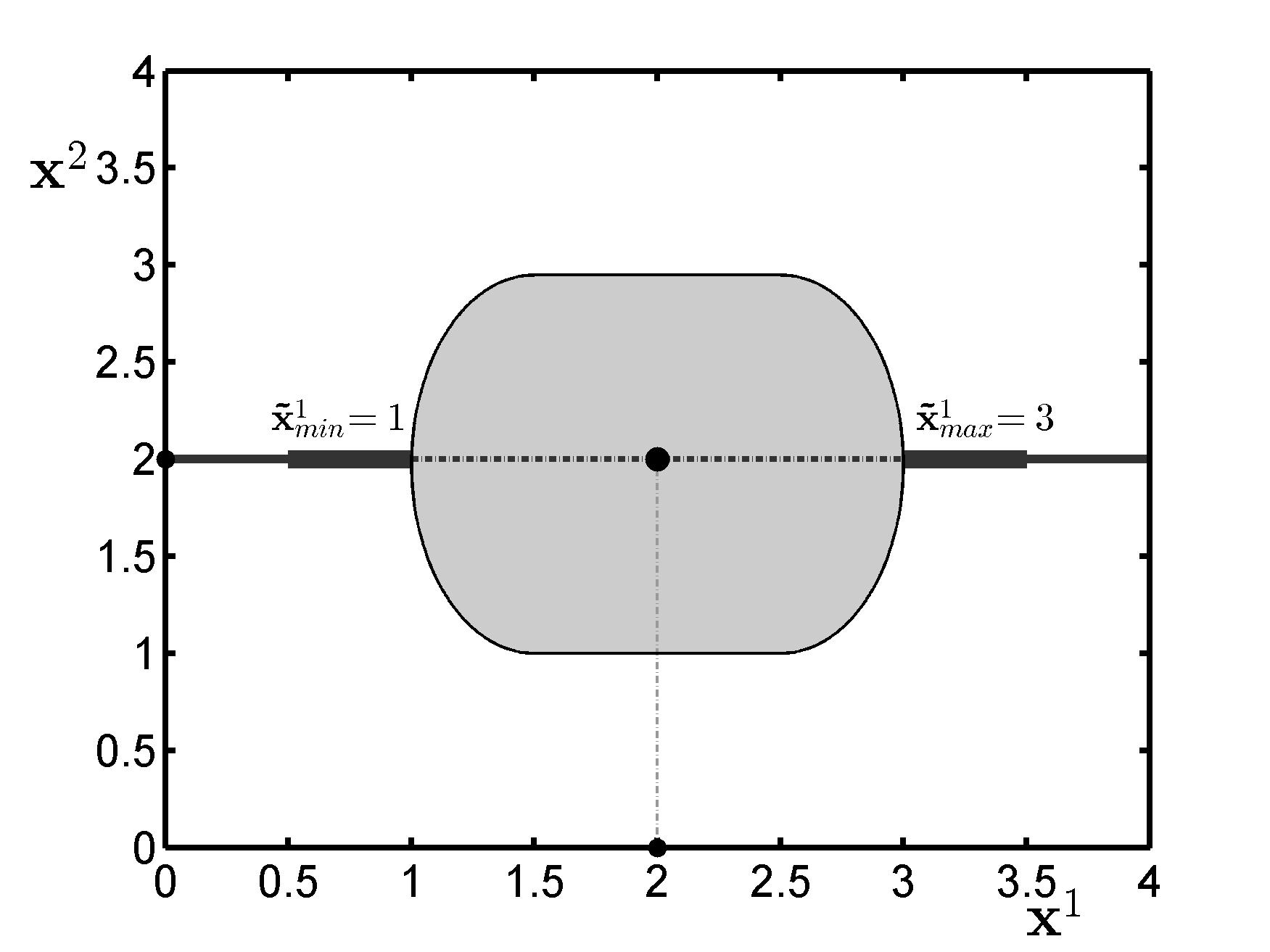}
\caption{\label{fig:Figure_1} Schematic illustrating the concept of single-node BS. The region inside the box comprises the entire state space of the two-oscillator network.}
\end{center}
\end{figure}

Let us assume that the networked dynamical system of Eq.~(\ref{eq:DE_Network}) has a stable synchronized state. Further, initiating the system from such a synchronized state, perturbations to even a single node of the system can drive the entire network of oscillators to a desynchronized state. For example, consider the simplest case of two oscillators (each exhibiting one-dimensional nodal dynamics) represented by the state variables $\mathbf{x}^{1}(t)$ and $\mathbf{x}^{2}(t)$ such that the synchronized state is described by the fixed point, $\mathbf{\tilde{x}} = \left(\, \mathbf{\tilde{x}}^{1},\, \mathbf{\tilde{x}}^{2}\, \right) = \left(\, \mathbf{\tilde{x}}^{*},\, \mathbf{\tilde{x}}^{*}\, \right) = \left(\, 2,\, 2\, \right)$ as illustrated in Fig.~\ref{fig:Figure_1}. The grey region indicates the basin of attraction of the synchronized state. Let the subspace of the first oscillator be confined between $\mathbf{x}^{1}_{\text{min}} = 0$ and $\mathbf{x}^{1}_{\text{max}} = 4$. Initiating the system from the synchronized state, the dashed line (at $\mathbf{x}^{2} = 2$) in Fig.~\ref{fig:Figure_1} visualizes the set of perturbations to the first oscillator after which the network converges back to the synchronized state. On the other hand, the solid lines in Fig.~\ref{fig:Figure_1} indicates the set of perturbations to the first oscillator which drive the network to the desynchronized state. Initiating the coupled system from the synchronized state, let $\mathbf{\tilde{x}}^{1}_{\text{min}}$ and $\mathbf{\tilde{x}}^{1}_{\text{max}}$ be the minimum and maximum admissible perturbed states, respectively, of the first oscillator, for which the coupled system will return to the synchronized state. The single-node BS of the first oscillator is the fraction of the one-dimensional volume of the state space of the respective oscillator belonging to the basin of attraction of the synchronized state. In other words, it is the ratio between the length of the dashed line and the lengths of the solid and dashed lines combined, i.e., $\frac{\mathbf{\tilde{x}}^{1}_{\text{max}} - \mathbf{\tilde{x}}^{1}_{\text{min}}}{\mathbf{x}^{1}_{\text{max}} - \mathbf{x}^{1}_{\text{min}}} = \frac{3 - 1}{4 - 0} = \frac{1}{2}$.

Formally, the single-node BS of the $i\textsuperscript{th}$ oscillator is defined as the fraction of the $d$-dimensional volume of the state space of the oscillator belonging to the $\left( d \times N \right)$-dimensional basin of attraction of the synchronized state. In the example presented in Fig.~\ref{fig:Figure_1}, $d = 1$ and $N = 2$. Thus, the single-node BS of any particular node of the network measures the probability of the system to remain in the basin of attraction of the synchronized state when random perturbations affect only that specific node.

Now, we present details on calculating the single-node BS value of the $i\textsuperscript{th}$ oscillator/node of a network (modeled using Eq.~(\ref{eq:DE_Network})). Perturbations to a (networked) dynamical system (and its nodes) are practically confined to a part of the state space which we refer to as the reference subset, in accordance with the terminology used by Menck et al.~\cite{menck2013basin}. In the computation of single-node BS of a particular node of the network, perturbations to that specific node are realized by giving the respective oscillator initial conditions chosen randomly from the reference subset, while initiating the other oscillators from the synchronized state. For example, in the illustration given in Fig.~\ref{fig:Figure_1}, we arbitrarily consider that the dashed line and the thick solid lines (at $\mathbf{x}^{2} = 2$) comprise the reference subset of the first oscillator. Put simply, these are the set of all possible initial conditions of the two-node system which we shall use for calculating the single-node BS of the first oscillator. Therefore, as a first step, select a reference subset $q$ of the phase space of the $i\textsuperscript{th}$ oscillator. Thus, for a network of identical oscillators, $\mathcal{Q} \equiv q^{N}$ comprises the reference subset of the complete $\left( d \times N \right)$-dimensional dynamical system. Below, we present the algorithm for calculating single-node BS:
\begin{enumerate}[label=(\roman*)]
\item Calculate the synchronization manifold $\mathbf{\tilde{x}(t)} = \left( \mathbf{\tilde{x}}^{1},\, \mathbf{\tilde{x}}^{2},\, \ldots,\, \mathbf{\tilde{x}}^{N} \right)^{\text{T}}$.
\item When the attractor corresponding to the synchronized state is not a fixed point, choose $P\ \left( > 1 \right)$ different points on the synchronization manifold. Otherwise, choose $P = 1$. In the former setting, the value of $P$ as well as the $P$ different points on the synchronization manifold have to be chosen such that these points sufficiently trace all parts of the attractor corresponding to the synchronized state.
\item For a particular value of $p\, \left( p = 1,\, 2,\, \ldots, \, P \right)$, perturb the $i\textsuperscript{th}$ oscillator by uniformly drawing $I_{C}$ random initial conditions from $q$, while each time initiating the system from the synchronized state using the $p\textsuperscript{th}$ point on the synchronization manifold.
\item Count the number $F_{C}$ of initial conditions that arrive at the synchronized state and estimate single-node BS of the $i\textsuperscript{th}$ oscillator $\left( S_{B}^{1} \left( i,\, p \right) \right)$ for the $p\textsuperscript{th}$ point on the synchronization manifold as 
\begin{equation}
\hat{S}_{B}^{1} \left( i,\, p \right) = \frac{F_{C}}{I_{C}}.
\end{equation}
\item Finally, average over $p$ to obtain the (mean) single-node BS value of node $i$,
\begin{equation} \label{eq:SNBS}
\langle S_{B}^{1} \left( i \right) \rangle = \frac{1}{P} \sum\limits_{p = 1}^{P} \hat{S}_{B}^{1} \left( i,\, p \right).
\end{equation}
\end{enumerate}

The concept of single-node BS is appropriate for extracting the contributions of individual nodes to the overall stability of the synchronized state. Further, it can be utilized to identify particularly vulnerable nodes of the system as well as more resilient ones.

By additionally averaging Eq.~(\ref{eq:SNBS}) over all nodes $i$, we may obtain a mean single-node BS value for the network as a whole, denoted as $\langle S_{B}^{1} \rangle$. Note that this property is distinctively different from the ``global'' BS $S_{B}$ of Menck et al.~\cite{menck2013basin} as it represents average information related to localized perturbations instead of such affecting the whole network at the same time. In this respect, this distinction is similar to that between global clustering coefficient (average local property) and network transitivity (global property) in the structural characterization of complex networks~\cite{radebach2013disentangling, donges2012analytical}.


\subsection{\label{sec:MNBS}Multi-node basin stability}

Now we consider $m \left( \geq 1 \right)$ nodes of the network being simultaneously perturbed such that the individual perturbations are independent of each other. In the following, we present details on calculating the multi-node BS, hereafter also referred to as \emph{m-node BS} value of the network.
\begin{enumerate}[label=(\roman*)]
\item For any particular value of $m$, generate an ensemble $\{E_j^m\}$ of \emph{$m$-node sets}, each consisting of $m$ nodes to be simultaneously perturbed. For an $N$-node network, there exist a possible maximum of ${N \choose m}$ of such $m$-node sets. At this point, multi-node BS can also be utilized for probing the influence of spatially localised perturbations or targeted attacks to specific parts of a network by selecting a \emph{specific} $m$-node set or a small ensemble thereof. For instance, given a spatially embedded network, one could perturb $m$ nodes from a localised region. In this paper, for any particular value of $m$, we randomly choose $M$ or ${N \choose m}$ (whichever is less) $m$-node sets, and leave the explicit investigation of different perturbation configurations as a subject of future research.
\item Given a particular $j\textsuperscript{th}$ $m$-node set $E_{j}^{m}$ of the ensemble, for any particular value of $p$, collectively perturb the $m$ nodes by uniformly drawing $I_{C}$ random initial conditions from $q^{m}$, while each time initiating the system from the $p\textsuperscript{th}$ point on the synchronization manifold (Section \ref{sec:SNBS}).
\item Count again the number $F_{C}$ of initial conditions that arrive at the synchronized state and estimate the $m$-node BS $S_{B}^{m} \left( E_{j},\, p \right)$ of the $j\textsuperscript{th}$ $m$-node set of the ensemble for the $p\textsuperscript{th}$ point on the synchronization manifold as 
\begin{equation}
\hat{S}_{B}^{m} \left( E_{j},\, p \right) = \frac{F_{C}}{I_{C}}.
\end{equation}
\item Finally, average over $p$ as well as over all the $m$-node sets of the ensemble to obtain the (mean) $m$-node BS value of the network as,
\begin{equation}
\langle S_{B}^{m} \rangle = \frac{1}{\text{min}\left( M,\, {N \choose m} \right)} \sum\limits_{j} \frac{1}{P} \sum\limits_{p} \hat{S}_{B}^{m} \left( E_{j},\, p \right).
\end{equation}
\end{enumerate}

For $m=1$, we obtain $\langle S_{B}^1 \rangle$ as described above as a special case.

The total number of subsets of nodes $\left( {N \choose m} \right)$ which can be simultaneously perturbed is generally very large making it computationally extremely expensive to compute multi-node BS. Therefore, to obtain a computationally feasible estimate of multi-node BS, we consider a smaller number $M\ \left( M \ll {N \choose m} \right)$ of such $m$-node sets selected uniformly at random~\cite{note_fnode_combinations}.

The framework of multi-node BS is highly relevant for assessing global stability and robustness of networked dynamical systems in response to non-local perturbations simultaneously affecting multiple nodes. Importantly, it provides an estimate of the critical number of nodes $\left( m_{crit} \right)$ which when simultaneously perturbed significantly reduces the ability of the system to return to the desired stable state. For the illustrations provided in the remainder of this paper (Section \ref{sec:Examples}), we estimate $m_{crit}$ by setting a threshold value of multi-node BS $\left( \langle S_{B} \rangle_{th} \right)$, such that $m_{crit}$ is defined as the minimum value of $m$ for which $\langle S_{B}^{m} \rangle \leq \langle S_{B} \rangle_{th}$. The value of $\langle S_{B}^{m} \rangle$ generally falls drastically with increase in the sizes of the reference subsets of the individual nodes and can be potentially small for large reference subsets~\cite{note_fnode_BS_threshold_f_crit}.


\section{\label{sec:Examples}Examples}


\subsection{\label{sec:DSFN_RO}Deterministic scale-free network of R\"{o}ssler oscillators}

\begin{figure*}
\begin{center}
\includegraphics[height=8.0cm, width=18.0cm]{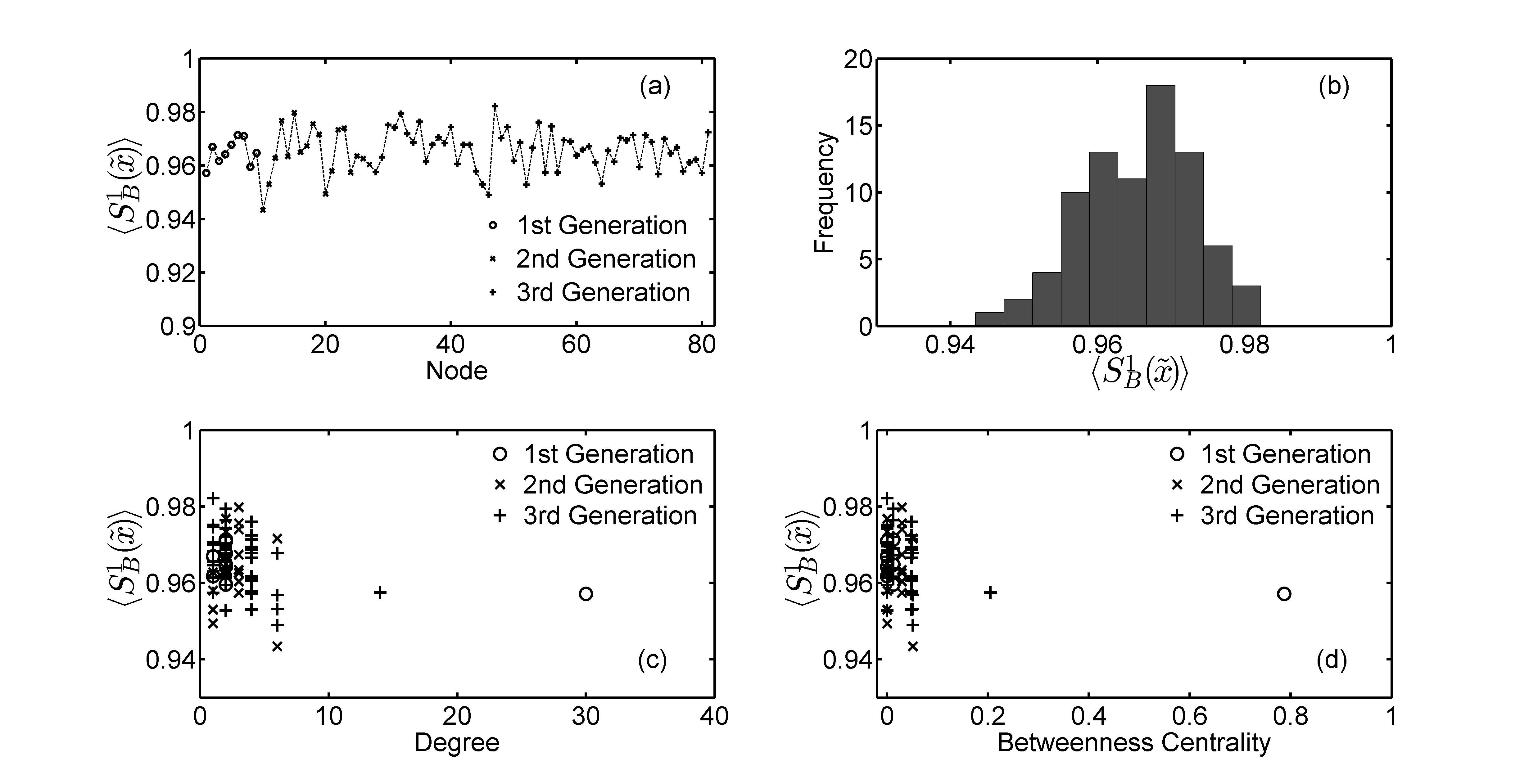}
\caption{\label{fig:Figure_2} (a) (Mean) single-node BS $\langle S_{B}^{1} \rangle$ of all the $N = 81$ nodes of the 3 generations of the undirected deterministic scale-free network of $N$ identical R\"{o}ssler oscillators. The first 9 nodes comprise the $1\textsuperscript{st}$ generation, the next 18 nodes the $2\textsuperscript{nd}$ generation and the final 54 nodes the $3\textsuperscript{rd}$ generation. (b) Histogram of $\langle S_{B}^{1} \rangle$ of all the $N = 81$ nodes. (c, d) Relationship of $\langle S_{B}^{1} \rangle$ with (c) degree and (d) betweenness centrality of the nodes, respectively.}
\end{center}
\end{figure*}

\begin{figure*}
\begin{center}
\includegraphics[height=7.0cm, width=18.0cm]{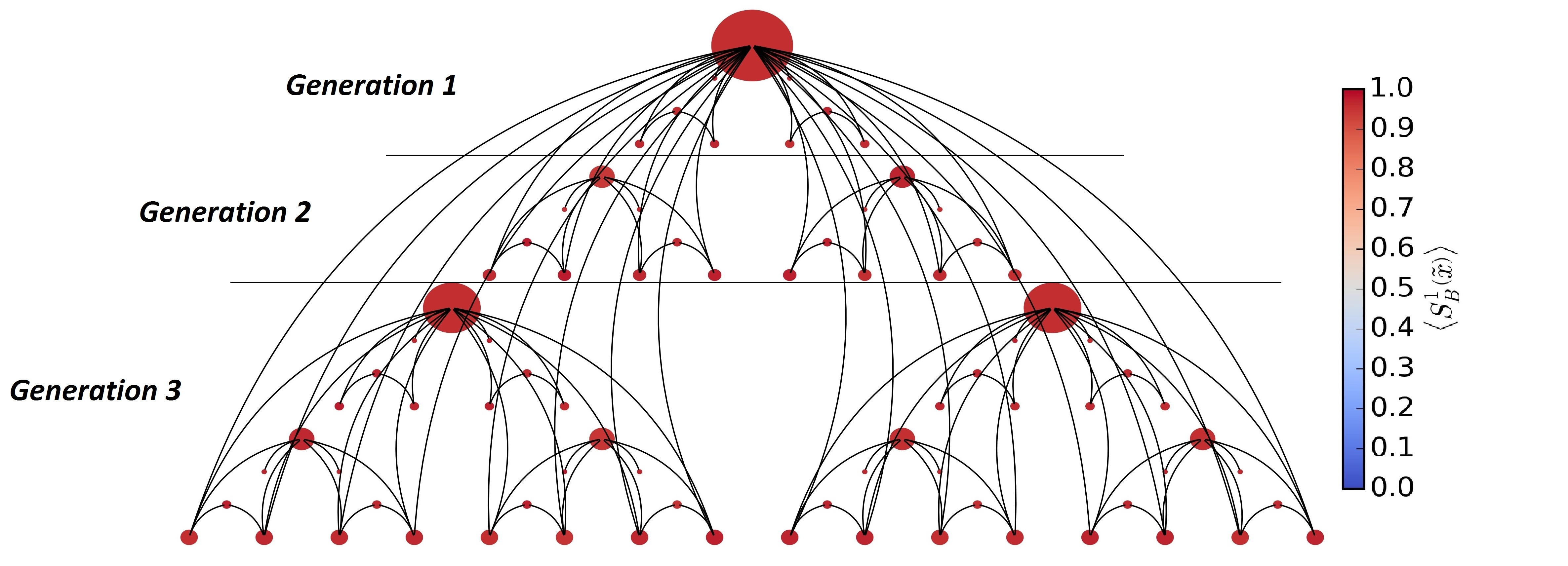}
\caption{\label{fig:Figure_3}(Color online) Network topology of the undirected deterministic scale-free network of $N = 81$ identical R\"{o}ssler oscillators. The size of each node is proportional to the degree and the color indicates the $\langle S_{B}^{1} \rangle$ value of the respective node.}
\end{center}
\end{figure*}

\begin{figure}
\begin{center}
\includegraphics[height=7.0cm, width=9.0cm]{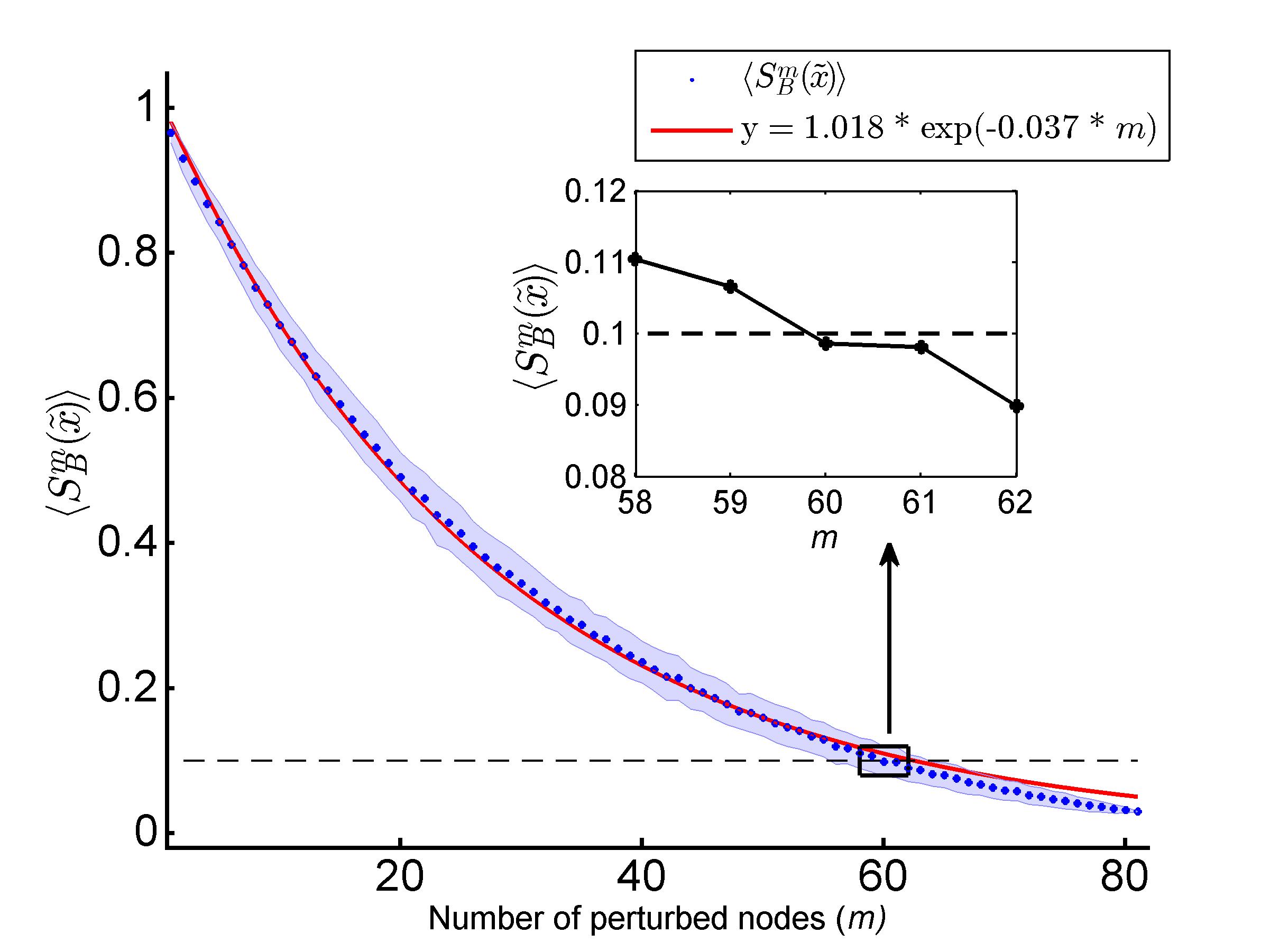}
\caption{\label{fig:Figure_4}(Color online) (Mean) $m$-node BS $\langle S_{B}^{m} \rangle$ (blue dots) for $m$ varying from $1$ to $N = 81$ in the undirected deterministic scale-free network of $N$ identical R\"{o}ssler oscillators. The shaded areas are representative of the standard deviations of the $m$-node BS values for the ensembles of \emph{$m$-node sets} chosen for computing $\langle S_{B}^{m} \rangle$ for a particular value of $m$. The red line is an exponential fit of $\langle S_{B}^{m} \rangle\ \left( \approx 1.018 * \exp \left( -0.037 * m \right) \right)$.}
\end{center}
\end{figure}

We consider a network of $N$ identical R\"{o}ssler oscillators, where the autonomous evolution of each individual unit is given by~\cite{rossler1976equation}
\begin{equation} \label{eq:RO_DE}
\begin{split}
\dot{x}_{1} & = - x_{2} - x_{3},\\
\dot{x}_{2} & = x_{1} + a x_{2},\\
\dot{x}_{3} & = b + x_{3} \left( x_{1} - c \right).
\end{split}
\end{equation}
We use the parameter values of $a = b = 0.2$ and $c = 7.0$ for which each uncoupled R\"{o}ssler oscillator in Eq.~(\ref{eq:RO_DE}) exhibits chaotic dynamics. We consider diffusive coupling in the $y$-variable between two coupled nodes such that the full dynamical equations of node $i$ (in analogy with Eq.~(\ref{eq:DE_Network})) read
\begin{equation} \label{eq:DSFN_RO_DE}
\begin{split}
\dot{x}_{1}^{i} & = - x_{2}^{i} - x_{3}^{i},\\
\dot{x}_{2}^{i} & = x_{1}^{i} + a x_{2}^{i} + \epsilon \sum\limits_{j = 1}^{N} A_{ij} \left( x_{2}^{j} - x_{2}^{i} \right),\\
\dot{x}_{3}^{i} & = b + x_{3}^{i} \left( x_{1}^{i} - c \right).
\end{split}
\end{equation}
We consider an undirected deterministic scale-free topology proposed by Barab\'{a}si, Ravasz and Vicsek~\cite{barabasi2001deterministic} and studied analytically by Iguchi and Yamada~\cite{iguchi2005exactly}. Such networks characterized by their fractal growth fall into the general class of hierarchical networks~\cite{ravasz2003hierarchical}. For the simulations carried out in this paper, we generate a deterministic scale-free network developed over 3 generations comprising $N = 81$ nodes.

\subsubsection*{\label{sec:DSFN_RO_SNBS}Single-node basin stability}

We are interested in the stability of the completely synchronized state, which corresponds to all oscillators following the same trajectory. In this context, we select a reference subset for each node as $q = [-15,\, 15] \times [-15,\, 15] \times [-5,\, 35]$; $\epsilon = 0.8$ is chosen from the stability interval predicted by the MSF~\cite{pecora1998master}; $P = 10$ points on the attractor of the completely synchronized state and $I_{C} = 500$ trials for estimating the (mean) single-node BS $\left( \langle S_{B}^{1} \rangle \right)$ values, using the procedure described in Section \ref{sec:SNBS}.

We calculate and present the $\langle S_{B}^{1} \rangle$ values of all the $N = 81$ nodes in Fig.~\ref{fig:Figure_2}(a). Interestingly, all the nodes have similar and relatively high $\langle S_{B}^{1} \rangle$ values (as also evident from the histogram in Fig.~\ref{fig:Figure_2}(b)). Figure~\ref{fig:Figure_2}(c, d) shows the relationship of the $\langle S_{B}^{1} \rangle$ values with the topological features of degree and betweenness centrality of the nodes, respectively. Apparently, the $\langle S_{B}^{1} \rangle$ values within any particular generation do not show a strong trend with respect to the degree or betweenness centrality of the nodes. This is further validated by the cross-correlation values of -0.2687 and -0.2108 of $\langle S_{B}^{1} \rangle$ with degree and betweenness centrality, respectively. We summarize our results in Fig.~\ref{fig:Figure_3}, which displays the network topology where the size of each node is proportional to the degree and the color corresponds to the $\langle S_{B}^{1} \rangle$ value of the respective node.

\subsubsection*{\label{sec:DSFN_RO_MNBS}Multi-node basin stability}

Next, using the algorithm described in Section \ref{sec:MNBS}, we calculate the (mean) $m$-node BS $\left( \langle S_{B}^{m} \rangle \right)$ values (for $M = 200$) and show the results in Fig.~\ref{fig:Figure_4} for $m$ varying from $1$ to $N \left( = 81 \right)$. Clearly, $\langle S_{B}^{m} \rangle$ declines significantly with increasing $m$. Interestingly, we observe that the variation of $\langle S_{B}^{m} \rangle$ values with increasing $m$ can be suitably modeled by an exponentially decaying function as illustrated in Fig.~\ref{fig:Figure_4}. Future studies on vulnerability of networked dynamical systems should focus on investigating and unraveling the mechanism underlying this observation.

In our example, we set $\langle S_{B} \rangle_{th} = 0.1$ and find (from the inset in Fig.~\ref{fig:Figure_4}) that $\langle S_{B}^{m} \rangle < \langle S_{B} \rangle_{th}$ for $m \ge 60$, implying $m_{crit} = 60$. Thus, simultaneously perturbing more than 60 nodes of the network (on average) significantly reduces the stability of the synchronized state below the critical threshold of $\langle S_{B} \rangle_{th} = 0.1$. We emphasize that this value refers to the average response of the network to randomly located perturbations. In case of targeted attacks, a much lower number of affected nodes can be sufficient to drive the system out of the synchronized state. In order to further address this aspect, it would be worth considering the full distribution of individual multi-node BS values $\hat{S}_B^m(E_j,p)$ for all $m$-node subsets and use the associated minimum/maximum values for describing worst-/best-case situations. We leave a detailed exploration of this problem as a subject for future work.

Our example clearly illustrates how $m$-node BS turns out to be a relevant concept for gauging the vulnerability of networked dynamical systems to global perturbations and emerges as a useful measure of the minimum fraction of nodes (on average) which (when perturbed simultaneously) significantly reduces the stability of the synchronized state. In turn, we recommend controlling or safeguarding at least $N - 60 = 21$ nodes of the network (on average) to ensure its functionality in the synchronized state in the face of large perturbations. Depending on the choice of the critical threshold $\langle S_{B} \rangle_{th}$, the value of $m_{crit}$ and, hence, the number of nodes to be controlled will vary. Notably, as for the detailed investigation of ``attack efficiencies'' of different $m$-node subsets discussed above, the problem of ``optimal control'' for safeguarding the synchronized state would also require further investigation of the full distribution of individual multi-node BS values $\hat{S}_B^m(E_j,p)$.


\subsection{\label{sec:PG_SOKM_UK}Power grid of the United Kingdom}

\begin{figure}
\begin{center}
\includegraphics[height=9.0cm, width=7.5cm]{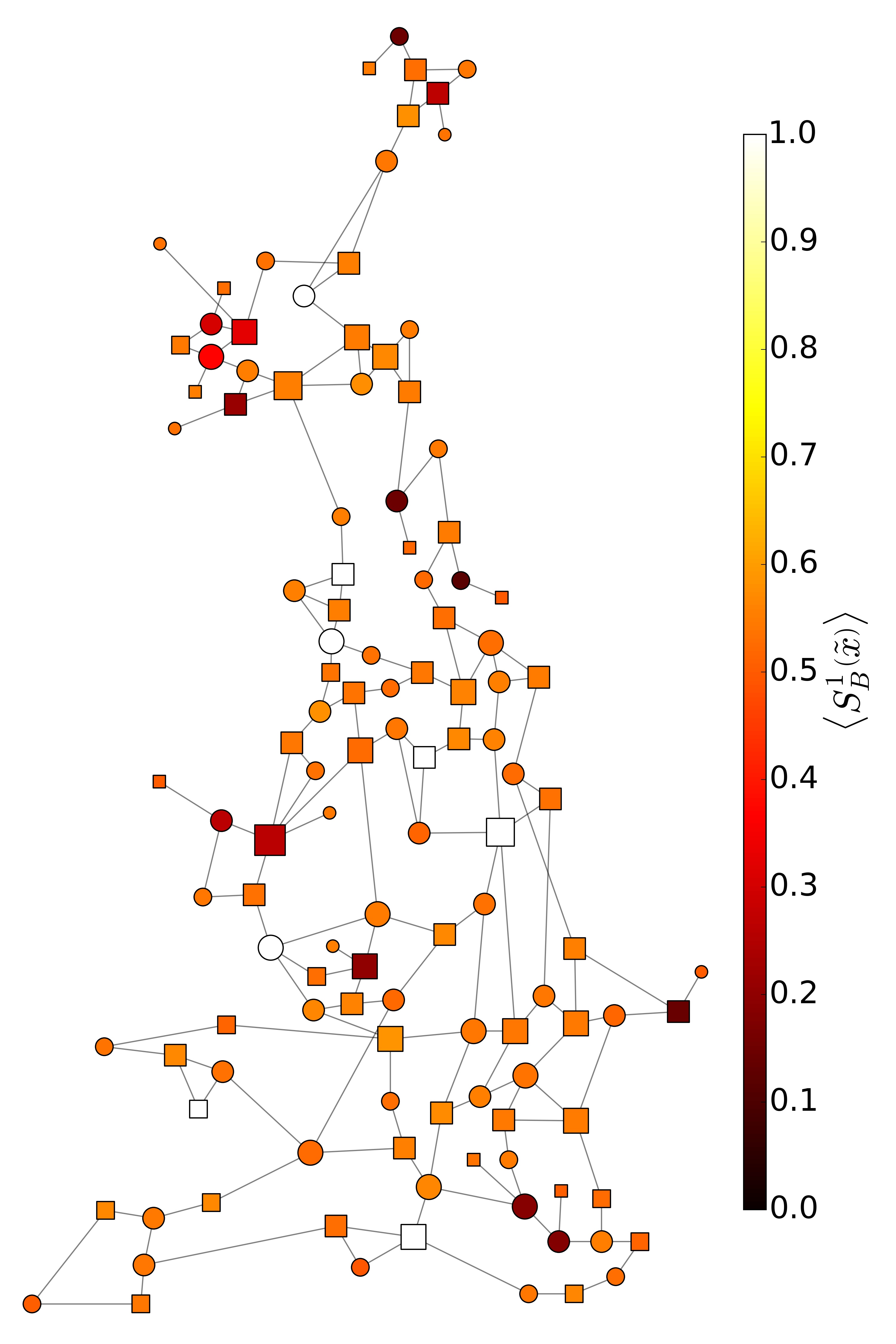}
\caption{\label{fig:Figure_5}(Color online) Network topology of the power transmission grid of the United Kingdom (comprising $N = 120$ nodes) with second-order Kuramoto-type nodal dynamics. Circular nodes are net generators while square nodes are net consumers. The size of each node is proportional to the degree, and its color corresponds to the $\langle S_{B}^{1} \rangle$ value of the respective node.}
\end{center}
\end{figure}

\begin{figure*}
\begin{center}
\includegraphics[height=8.0cm, width=18.0cm]{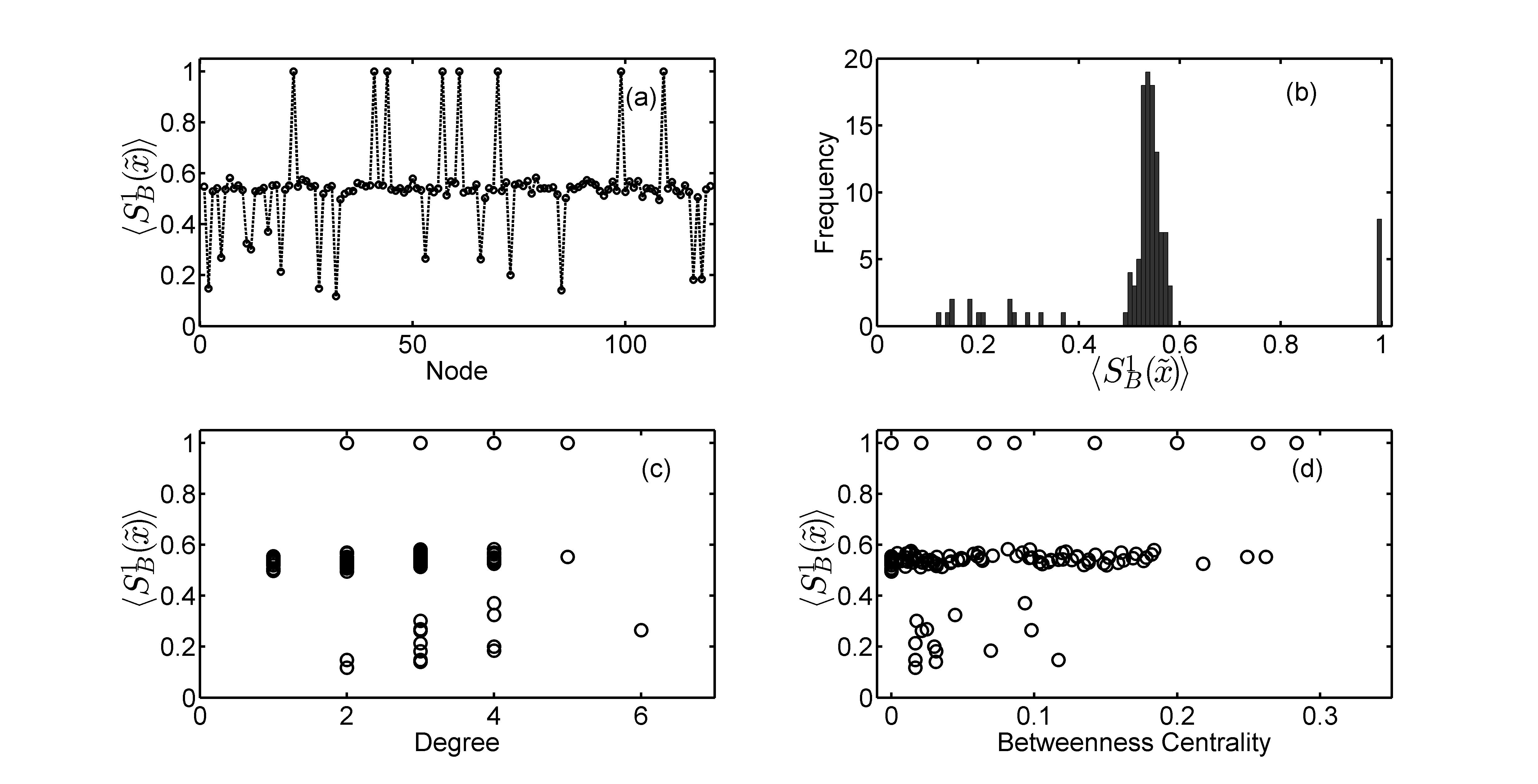}
\caption{\label{fig:Figure_6} As in Fig.~\ref{fig:Figure_2} for the $N = 120$ nodes of the power grid of the United Kingdom with second-order Kuramoto-type nodal dynamics.}
\end{center}
\end{figure*}

\begin{figure}
\begin{center}
\includegraphics[height=7.0cm, width=9.0cm]{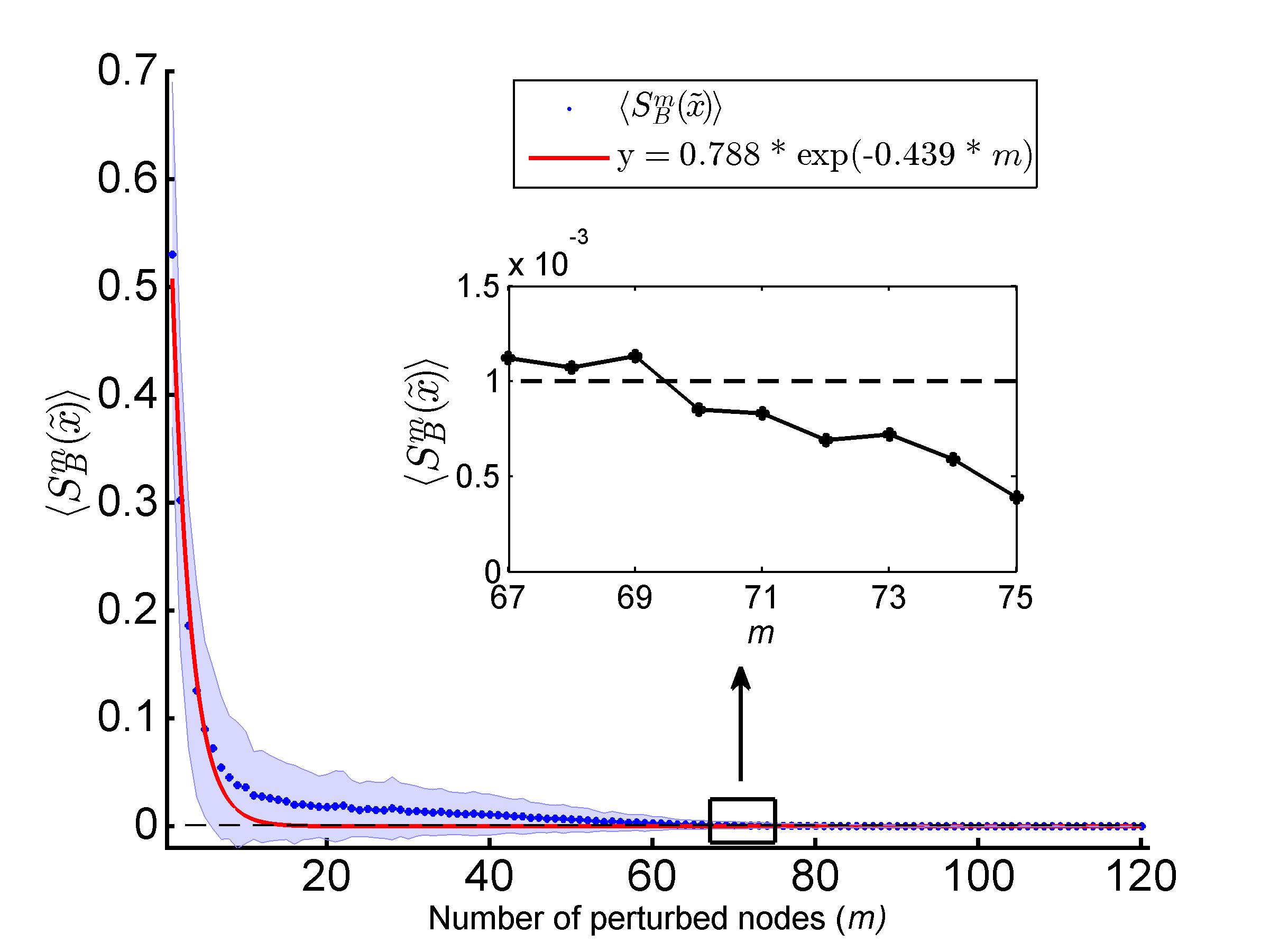}
\caption{\label{fig:Figure_7}(Color online) (Mean) $m$-node BS $\langle S_{B}^{m} \rangle$ (blue dots) for $m$ varying from $1$ to $N = 120$ for the power grid of the United Kingdom with second-order Kuramoto-type nodal dynamics. The shaded background is representative of the standard deviation of the $m$-node BS values for the ensemble of \emph{$m$-node sets} chosen for computing $\langle S_{B}^{m} \rangle$ for a particular value of $m$. The red line shows an exponential fit of $\langle S_{B}^{m} \rangle$.}
\end{center}
\end{figure}

As a more realistic example, we consider a conceptual model of the power transmission grid of the United Kingdom with second-order Kuramoto-type nodal dynamics~\cite{rohden2012self, witthaut2012braess, rodrigues2016kuramoto}. The network consists of $N = 120$ nodes and 165 transmission lines (as illustrated in Fig.~\ref{fig:Figure_5}) which corresponds to a mean nodal degree of 2.75~\cite{manik2014supply}. 

In order to capture the relevant dynamical aspects and collective phenomena exhibited by a power grid, we consider a coarse-scale model comprising second-order Kuramoto-type oscillators coupled on the aforementioned network topology~\cite{rohden2012self, witthaut2012braess, rodrigues2016kuramoto}. Such models consist of synchronous generators (representing power plants) and motors (representing consumers) characterized by the electrical power $P_{i}$ the machines generate ($P_{i} > 0$) and consume ($P_{i} < 0$), respectively. The dynamical state of each machine is represented by its mechanical phase $\phi_{i}(t) = \Omega t + \theta_{i}(t)$ and its phase velocity $\dot{\phi}_{i}(t)$, where $\Omega \left( = 2 \pi\, \times\, 50\ \text{Hz}\ \text{or}\ 2 \pi\, \times\, 60\ \text{Hz}\right)$ is the reference frequency of the grid. Considering the law of conservation of energy, the power generated or consumed by each unit $P_{source,\ i}$ must be equal to the sum of its power exchanged (given to or taken from) with the grid $P_{trans,\ i}$, its accumulated power $P_{acc,\ i}$ and its dissipated power $P_{diss,\ i}$. The power dissipated by each machine is given by $P_{diss,\ i} = \kappa_{i} \dot{\phi_{i}}^{2}$ where $\kappa_{i}$ is the dissipation coefficient of the respective unit. The power accumulated by each rotating machine is given by $P_{acc,\ i} = \frac{dE_{kin,\ i}}{dt}$ where $E_{kin,\ i} = \frac{I \dot{\phi_{i}}^{2}}{2}$ is the kinetic energy and $I_{i}$ is the moment of inertia of the respective unit. The power transmitted between two machines $i$ and $j$ is given by $P_{trans,\ ij} = P_{max,\ ij}\, \sin \left( \phi_{i} - \phi_{j} \right)$, where $P_{max,\ ij}$ is the maximum capacity of the respective transmission line. The condition of conservation of energy at each node of the network yields
\begin{equation} \label{eq:PG_SOKM_DE_1}
\begin{split}
P_{source,\ i} & = P_{diss,\ i} + P_{acc,\ i} + P_{trans,\ i},\\
& = P_{diss,\ i} + P_{acc,\ i} + \sum\limits_{j = 1}^{N} P_{trans,\ ij},\\
& = \kappa_{i} \dot{\phi_{i}}^{2} + I \ddot{\phi_{i}} + \sum\limits_{j = 1}^{N} P_{max,\ ij}\, \sin \left( \phi_{i} - \phi_{j} \right).\\
\end{split}
\end{equation}
For simplicity, we consider that all the machines have identical moment of inertia $I_1 = \ldots = I_N = I$ and dissipation coefficient $\kappa_1 = \ldots \kappa_N = \kappa$. Further, substituting $\phi_{i}(t) = \Omega t + \theta_{i}(t)$ in Eq.~(\ref{eq:PG_SOKM_DE_1}) and assuming that the phase changes are much slower than the reference frequency (i.e. $\arrowvert \dot{\theta} \arrowvert \ll \Omega$) leads to the following equation of motion:
\begin{equation} \label{eq:PG_SOKM_DE_2}
\ddot{\theta}_{i} = -\alpha \dot{\theta}_{i} + P_{i} + \sum\limits_{j = 1}^{N} \epsilon_{ij} \sin \left( \theta_{i} - \theta_{j} \right),
\end{equation}
where $\alpha = \frac{2 \kappa}{I}$, $P_{i} = \frac{P_{source,\ i} - \kappa \Omega^{2}}{I \Omega}$ and $\epsilon_{ij} = \frac{P_{max,\ ij}}{I \Omega}$. We further assume the capacity of all transmission lines to be equal, i.e. $\epsilon_{ij} = \epsilon A_{ij}$ (where $\epsilon$ and $\mathbf{A}$ again denote the overall coupling strength and the adjacency matrix, respectively), such that the dynamical equations of the system (in analogy with Eq.~(\ref{eq:DE_Network})) read
\begin{equation} \label{eq:PG_SOKM_UK_DE}
\begin{split}
\dot{\theta}_{i} & = \omega_{i},\\
\dot{\omega}_{i} & = - \alpha \omega_{i} + P_{i} + \epsilon \sum\limits_{j = 1}^{N} A_{ij} \sin \left( \theta_{j} - \theta_{i} \right),
\end{split}
\end{equation}
where $\omega_{i}$ denotes the frequency of the $i\textsuperscript{th}$ oscillator. Furthermore, we randomly choose $\frac{N}{2}$ net generators and $\frac{N}{2}$ net consumers with $P_{i} = +P_{0}$ and $P_{i} = -P_{0}$, respectively~\cite{menck2014dead}. In the following, we use the parameter values $\alpha = 0.1$, $P_{0} = 1.0$ and $\epsilon = 8.0$ for obtaining the results described below.

\subsubsection*{\label{sec:PG_SOKM_UK_SNBS}Single-node basin stability}

We consider the stability of the synchronized state, which corresponds to all oscillators having constant phases $\tilde{\theta^{i}}$ and frequencies $\tilde{\omega^{i}} = 0$. We select a reference subset for each node as $q = [0,\, 2\pi] \times [-100,\, 100]$; $P = 1$ point on the attractor of the synchronized state and $I_{C} = 1000$ trials. The $\langle S_{B}^{1} \rangle$ values of all the $N = 120$ nodes are shown in Fig.~\ref{fig:Figure_6}(a). Figure~\ref{fig:Figure_6}(b) displays a histogram of all $\langle S_{B}^{1} \rangle$ values, where the nodes split into three classes displaying poor $\left( \langle S_{B}^{1} \rangle \leq 0.4 \right)$, fair $\left( 0.4 < \langle S_{B}^{1} \rangle < 0.75 \right)$ and high $\left( \langle S_{B}^{1} \rangle \geq 0.75 \right)$ values of (mean) single-node BS. Figure~\ref{fig:Figure_6}(c, d) again illustrates the distribution of $\langle S_{B}^{1} \rangle$ in comparison with degree and betweenness centrality, respectively. The cross-correlation values of $\langle S_{B}^{1} \rangle$ with degree and betweenness centrality are 0.061 and 0.281, respectively, ruling out the existence of a systematic dependence between $\langle S_{B}^{1} \rangle$ and the two considered topological node characteristics. Figure~\ref{fig:Figure_5} displays the network topology together with the individual $\langle S_{B}^{1} \rangle$ values in full analogy with Fig.~\ref{fig:Figure_3} for the R\"{o}ssler network. Finally, note that the nodes which show up with the lowest $\langle S_{B}^{1} \rangle$ values comprise the dead ends of the network, in agreement with Menck et al.~\cite{menck2014dead}.

\subsubsection*{\label{sec:PG_SOKM_UK_MNBS}Multi-node basin stability}

Finally, we study the variation of the (mean) $m$-node BS $\langle S_{B}^{m} \rangle$ for $m$ again varying from $1$ to $N \left( = 120 \right)$ and $M = 1000$ or ${N \choose m}$ (whichever is less) randomly chosen $m$-node sets (Fig.~\ref{fig:Figure_7}). Clearly, $\langle S_{B}^{m} \rangle$ declines rapidly with increasing $m$ until $m \approx 12$ beyond which it decreases gradually until $m \approx 80$, thereafter saturating at a value $\approx 0.0005$. For $\langle S_{B} \rangle_{th} = 0.001$, we find (from the inset in Fig.~\ref{fig:Figure_7}) that $\langle S_{B}^{m} \rangle < \langle S_{B} \rangle_{th}$ for $m \ge 70$, i.e. $m_{crit} = 70$. Thus, safeguarding at least $50$ nodes of the network (on average) ensures functionality of the power grid in the synchronized state in the considered setting. Notably, we again observe that the decay of $\langle S_{B}^{m} \rangle$ values can be fitted by an exponential function of $m$ as illustrated in Fig.~\ref{fig:Figure_7}, suggesting that this feature is not exclusive to the hierarchical network organization of our first example.


\section{\label{sec:Conclusion}Conclusion}

The ubiquity of multistability in complex networks of dynamical systems calls for the development of suitable quantifiers of the respective stability of the multiple stable states of such systems. This has recently led to the development of basin stability (BS) and its extension to the concept of single-node BS. The single-node BS of a particular node of a network corresponds to the probability of the system to return to the desired stable state in the face of large perturbations hitting the respective node. However, in general, networked dynamical systems can also be subject to perturbations simultaneously affecting several nodes of the system. 

On this account, we proposed the general framework of \emph{multi-node basin stability} for gauging global stability and robustness of networked dynamical systems in response to non-local perturbations simultaneously hitting multiple nodes of the system. Although the established framework of master stability function (MSF) for assessing the stability of the synchronized state was a major advancement, it has still been locally confined to small perturbations. Moreover, the MSF-based approach is mostly restricted to studying the stability of synchronization in coupled identical (or nearly identical) systems~\cite{pecora1998master, sun2009master}. However, the framework of single-node BS and multi-node BS is applicable to non-identical systems as well as scenarios with non-identical functions coupling them. Importantly, multi-node BS provides an estimate of the minimum fraction of nodes (on average) which when perturbed simultaneously significantly hampers the ability of the system to return to the desired stable state. Furthermore, multi-node BS can also be used to identify the exact set of nodes/oscillators of the network which are most susceptible to perturbations and constitute the dynamically least robust sub-components of the network. As examples, we have studied the stability of the synchronized state in a deterministic scale-free network of R\"{o}ssler oscillators and a conceptual model of the United Kingdom power grid with second-order Kuramoto-type nodal dynamics. Subsequently, we foresee the general framework of multi-node BS as a paradigm for assessing stability and resilience in complex networks of dynamical systems from various fields of application.

Recently, Mitra et al.~\cite{mitra2015integrative} have reconsidered the concept of ecological resilience~\cite{walker2004resilience} in proposing the framework of \emph{integral stability} as a holistic quantifier of multistability. Immediate future studies in line with the concept of multi-node BS could constitute its extension to a framework of \emph{single-}and \emph{multi-node integral stability}. In the examples presented above, multi-node BS has been applied to networks of identical oscillators. Thus, another logical extension of the present work should constitute its application to probing multistability in networks of non-identical oscillators. Furthermore, multi-node BS can be applied to assessing the stability of interdependent networks of dynamical systems. Finally, we believe that the framework of multi-node BS can be applied to revealing the underlying structure of a complex network by examining the responses of different set of nodes/oscillators to localized perturbations.


\begin{acknowledgments}
CM and RVD have been supported by the German Federal Ministry of Education and Research (BMBF) via the Young Investigators Group CoSy-CC\textsuperscript{2} (grant no. 01LN1306A). AC acknowledges the JC Bose fellowship (SB/S2/JCB-013/2015) for financial support. The authors thank Paul Schultz for providing the data on the topology of the United Kingdom power grid.
\end{acknowledgments}


\bibliography{References}

\end{document}